\documentclass[aps,prl,twocolumn,superscriptaddress]{revtex4-2}
\usepackage{graphicx}
\begin{document}


\title{Microtearing Turbulence and Its Role in High-Density-Gradient Plasmas in Wendelstein 7-X}

\author{H. Cu-Castillo}
\email[]{hugo.cu.castillo@ipp.mpg.de}
\author{A. Bañón Navarro}
\author{G. Merlo}
\affiliation{Max-Planck-Institut für Plasmaphysik, Boltzmannstraße 2, 85748 Garching, Germany}
\author{F. Reimold}
\author{T. Romba}
\author{O. Ford}
\author{S. Bannmann}
\author{L. Vanó}
\author{M. Wappl}
\author{J. Geiger}
\author{A. Zocco}
\affiliation{Max-Planck-Institut für Plasmaphysik, Wendelsteinstraße 1, 17489 Greifswald, Germany}
\author{F. Jenko}
\affiliation{Max-Planck-Institut für Plasmaphysik, Boltzmannstraße 2, 85748 Garching, Germany}
\author{the W7-X team}
\affiliation{Max-Planck-Institut für Plasmaphysik, Wendelsteinstraße 1, 17489 Greifswald, Germany}

\date{\today}

\begin{abstract}
Gyrokinetic simulations reveal that microtearing mode (MTM) turbulence dominates transport in a Wendelstein 7-X (W7-X) discharge characterized by large density gradients, moderate temperature gradients, and low plasma beta. This conclusion is supported by the close agreement between simulated and experimentally measured heat and particle fluxes. The emergence of MTMs is attributed to the absence of competing instabilities -- such as ion temperature gradient modes and density-gradient-driven trapped-electron modes -- under these plasma conditions, together with the stabilizing influence of the W7-X max-$J$ magnetic configuration. Furthermore, moderate collisionality and low magnetic shear are found to facilitate MTM onset. These findings advance the understanding of high-density-gradient regimes, which are essential for achieving high-performance operation in W7-X plasmas.
\end{abstract}

\maketitle


Wendelstein 7-X (W7-X) is the world’s most advanced stellarator and aims to demonstrate that its magnetic confinement concept is a viable path toward a fusion power plant. Recently, W7-X has achieved record-breaking discharges with performance levels on par with leading tokamaks~\cite{Bannmann25}, and its improved confinement can be sustained over long durations~\cite{Dinklage2025}. This success is largely attributed to its optimized magnetic configuration, designed to minimize neoclassical transport losses, among other criteria~\cite{Beidler21}. As in tokamaks, however, plasma confinement in W7-X is ultimately limited by microturbulence~\cite{Klinger19,Pedersen22,Wilms24,AgapitoFernando25,Wilms25}.

High-performance scenarios are characterized by steep density gradients ($\nabla n$), typically driven by core fueling techniques such as frozen hydrogen pellet injection~\cite{Bozhenkov2020} or neutral beam injection (NBI)~\cite{Ford24}. Under these conditions, the plasma exhibits reduced turbulent heat transport~\cite{Bozhenkov2020,Ford24,Wappl25} and favorable confinement of particles (protium ions and electrons)~\cite{Bannmann24}, but it may also suffer from impurity accumulation~\cite{Romba23,Romba25}. Understanding the nature of turbulence in these regimes is therefore crucial to achieve a balance between improved energy and particle confinement and the expulsion of impurities. 

This letter demonstrates that microtearing mode (MTM) turbulence is dominant in a plasma discharge with a high $\nabla n$, similarly to the high-performance regimes. MTMs are small-scale electromagnetic instabilities that cause tearing and reconnection of magnetic field lines. They are driven by the electron temperature gradient and are stabilized by magnetic shear, a measure of magnetic field line bending~\cite{Gladd80,Drake80}. In slab geometry (i.e., in a homogeneous magnetic field), they are caused by an imbalance in the force balance parallel to the magnetic field due to a time-dependent electron thermal force, and collisions can be crucial for their destabilization~\cite{Drake77,Zocco15}. MTM turbulence has been previously identified as relevant in several tokamak experiments~\cite{Guttenfelder11,Jian19,Hatch21}. In this work, we show for the first time that MTM turbulence can also be dominant in stellarator plasmas by simulating microturbulence in W7-X using the gyrokinetic code GENE~\cite{Jenko2000}. Our simulations can reproduce the experimentally observed heat and particle fluxes by including key physical effects previously neglected in studies on these scenarios -- namely, collisionality and electromagnetic fluctuations~\cite{Xanthopoulos20,Proll22,Zocco24}. 


\textit{Experimental scenario and numerical setup.} The analyzed case corresponds to the purely NBI-heated phase of the high-mirror configuration discharge \#20181009.034 in W7-X. The plasma is started with electron cyclotron resonance heating (ECHR) and fueling via a neutral gas inlet. Both systems are turned off once the plasma has sufficient density, then the NBI deposits fuel and a total injected power of $P\approx3.7$ MW over a broad plasma volume. The central density and its gradient increase, while the electron and ion temperatures remain constant and roughly equal to each other throughout this interval of the discharge~\cite{Bannmann24,Ford24}. The time examined in this work is $t = 2.8$~s and unless stated otherwise, the radial location inspected is $\rho \equiv r_\mathrm{eff}/a =0.4$. The effective radius is defined as $r_\mathrm{eff}=\sqrt{A/\pi}$, where $A$ is the area enclosing the selected flux surface and $a=0.52$~m is the plasma minor radius.
The numerical plasma parameters used to symbolically represent the experiment at this position are: $a/L_{n_e} = 2.37$, $a/L_{T_e} = 1.54$, $a/L_{T_i} = 1.91$, $T_i = 1.1 \cdot T_e$, $\beta_e = 8\pi n_e T_e / B_\mathrm{ref}^2 = 5.2 \times 10^{-3}$ (in CGS units), and normalized collision frequency~\cite{Merz2009} $\nu_c = \pi \ln\Lambda q_e^4 n_e a / (2^{3/2} T_e^2) = 2.21 \times 10^{-3}$. Here, $a/L_\xi = -a/\xi \cdot (d\xi/d\rho)$ denotes the normalized scale length of a given quantity $\xi$, $B_\mathrm{ref}$ is the reference magnetic field strength and the subscripts \{e,i\} refer to the corresponding electron and main ion quantities. The magnetic shear at this radial position is $\hat{s} = \rho_0/q_0 \cdot (dq/d\rho) = -0.021$, where $q = 1/\iota$ is the safety factor (with $\iota$ being the rotational transform), and $\rho_0$ and $q_0 = -1.14$ are the reference radial position and safety factor, respectively.

Simulations were performed using the local version of the gyrokinetic code \textsc{GENE}, which evolves small-scale plasma fluctuations while incorporating electromagnetic effects and collisions. The latter are modeled using the linearized Landau-Boltzmann operator. The simulation coordinates are aligned to the background magnetic field, $\mathbf{B}_0 = \nabla x \times \nabla y$, where $x = \rho$ is the radial coordinate, and $y$ is the binormal coordinate given by $y = x_0/q_0 \cdot (q_0 \theta^* - \phi)$, with $\theta^*=z$ the PEST poloidal angle~\cite{Li16} and $\phi$ the toroidal angle. The velocity space of the perturbed distribution function $\delta f$ is discretized in terms of the velocity parallel to the magnetic field line $v_\parallel$ and the magnetic moment $\mu = m_i v_\perp^2 / (2B_0)$.

\begin{figure}[t]
\includegraphics{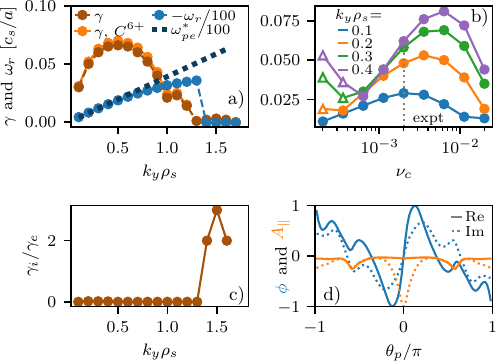}
\caption{\label{F1_lin_characterization} 
Linear gyrokinetic characterization of instabilities:  
a) Growth rate and real frequency as a function of $k_y \rho_s$. Impurity contribution (orange) is negligible. The dotted line indicates the electron diamagnetic frequency $\omega^*_{p_e}$.  
b) Growth rate versus normalized collision frequency for various $k_y \rho_s$. The experimental collisionality is marked by a vertical dashed line. Full circles denote MTM; hollow triangles denote a TPM. The changes for $k_y\rho_s=0.1$ were: $(n_x,n_z)=(64,288)$ and 3 npol.
c) Ratio of the contributions to the growth rate from ions and electrons.  
d) Tearing parity at $k_y \rho_s = 0.5$: odd parity in $\phi$ (blue) and even in $A_\parallel$ (orange), normalized to their respective maxima. Only the central ballooning angle region is shown.  
}
\end{figure}

We performed both linear and nonlinear simulations. Unless stated otherwise, for the linear runs, the numerical resolution was $(n_x, n_{k_y}, n_z) = (48, 1, 192)$ in the spatial directions, and $(n_{v_\parallel}, n_\mu) = (32, 12)$ in velocity space. The simulation domain in velocity space was $(l_{v_\parallel}, l_\mu) = (3\sqrt{2}c_s, 9T_e/B_0)$, where $c_s = \sqrt{T_e / m_i}$ is the reference speed. The standard twist-and-shift boundary condition~\cite{Beer95} was applied after two poloidal turns ("npol"), unless otherwise specified, in which case the generalized boundary condition of Ref.~\cite{Martin2018} was used. Nonlinear simulations employed a resolution of $(n_x, n_{k_y}, n_z, n_{v_\parallel}, n_\mu) = (288, 36, 768, 32, 24)$. The simulation domain sizes were $(l_x, l_y) = (288 \rho_s, 314.15 \rho_s)$, and eight poloidal turns. This study excludes impurity effects as their contribution has a negligible effect on the linear growth rates $\gamma$, as indicated in Fig. \ref{F1_lin_characterization}a. Experimentally, the impurity presence was characterized by $a/L_{n_{C^{6+}}} = 2.3$ and $n_{C^{6+}} = 3 \times 10^{-3} \cdot n_e$. Parallel magnetic field fluctuations were also excluded for the same reason (not shown).


\textit{Mode identification.} We performed a linear gyrokinetic stability analysis under the experimental conditions described above. The results, summarized in Fig.~\ref{F1_lin_characterization}, indicate that the dominant instability over a broad range of binormal wavenumbers ($0.1 \leq k_y \rho_s \leq 1.4$) is the microtearing mode. This is characterized by a negative real frequency $\omega_r < 0$, indicating propagation in the electron diamagnetic direction, as shown in Fig.~\ref{F1_lin_characterization}a. The reference gyroradius is $\rho_s = c_s / \Omega_\mathrm{ref}$, and $\Omega_\mathrm{ref} = q_e B_\mathrm{ref} / (m_i c)$ is the reference gyrofrequency (in CGS units). Moreover, the real frequency closely follows the electron diamagnetic drift frequency, $\omega_r \approx \omega^*_{p_e} = k_y \rho_s c_s (1/L_{T_e} + 1/L_n)$, consistent with the slab semi-collisional MTM regime predicted by analytical theory~\cite{Gladd80,Drake80}.

Additional features reinforce the identification of the linearly dominant mode as MTM. The parity structure of the eigenmodes (Fig.~\ref{F1_lin_characterization}d) has odd parity in the electrostatic potential $\phi$ and even parity in the parallel vector potential $A_\parallel$, as expected for tearing modes. The species-resolved contributions to the growth rate~\cite{Manas2015}, shown in Fig.~\ref{F1_lin_characterization}c shows a clear dominance of the electron response. Furthermore, the growth rate dependence on collisionality in Fig.~\ref{F1_lin_characterization}b exhibits a non-monotonic trend, a characteristic feature of MTM~\cite{Gladd80}, and the effect is stronger at larger wavenumbers, in agreement with previous tokamaks observations~\cite{Doerk12}. At lower collisionalities, a trapped-particle mode (TPM) emerges, characterized by its electrostatic nature with ballooning parity, $\omega_r \lesssim 0$, driven by $\nabla n$, and heat flux localization at magnetic wells (not shown), and is similar to what was referred to in Ref.~\cite{Plunk2017} as the ion-driven trapped-electron mode (iTEM). Because $\gamma_i/\gamma_e\gtrsim1$ for this case, it is unlikely to be the collisionless trapped-electron mode (TEM)~\cite{Proll13PoP}.

\begin{figure}[t]
\includegraphics{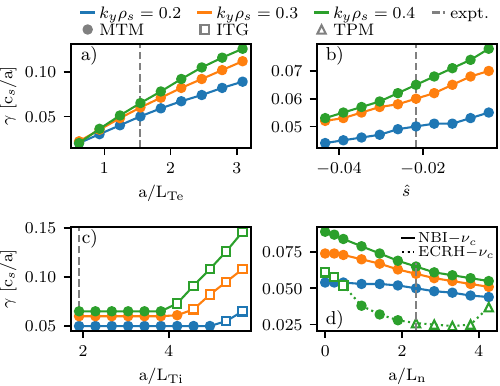}
\caption{\label{F4_shat_omTi_omTe_omn_scans} 
Dependence of MTM growth rate on:  
a) electron temperature gradient,  
b) global magnetic shear,  
c) ion temperature gradient, and  
d) density gradient.  
The nominal NBI collisionality (solid line) and the reduced collisionality of the ECRH reference case mentioned in the text were used (dotted line). Vertical dashed lines indicate experimental values.}
\end{figure}

\textit{Linear parametric scan.} The presence of the MTM instability in this NBI-heated plasma can be attributed to a combination of destabilizing and stabilizing mechanisms. According to slab semi-collisional theory, MTMs are driven by the electron temperature gradient and stabilized by magnetic field line bending, which scales with the magnetic shear $\hat{s}$~\cite{Gladd80,Drake80}. Although W7-X has a much more complex geometry than the slab model, we find that the qualitative behavior of MTMs remains consistent with these theoretical predictions. Fig.~\ref{F4_shat_omTi_omTe_omn_scans}a confirms that the MTM growth rate increases with the electron temperature gradient. Fig.~\ref{F4_shat_omTi_omTe_omn_scans}b shows that reducing $|\hat{s}|$ -- thereby weakening field line bending -- leads to enhanced MTM growth. In these scans, the parallel boundary condition was artificially modified to emulate the effect of varying the magnetic shear $\hat{s}$. Although this approach is not strictly self-consistent, it qualitatively isolates the influence of magnetic shear on the instability.

To further test the robustness of MTMs within experimental error bars, we scanned the ion temperature gradient. As shown in Fig.~\ref{F4_shat_omTi_omTe_omn_scans}c, its effect on MTMs is negligible except for $\eta_i = L_n / L_{T_i} \gtrsim 1.7$ -- about twice the experimental value $\eta_{i,\mathrm{expt}} = 0.8$ -- where ion temperature gradient (ITG) modes are found to be unstable.

MTM shows only weak sensitivity to the density gradient $a/L_n$ and is slightly stabilized as $a/L_n$ increases (Fig.~\ref{F4_shat_omTi_omTe_omn_scans}d). Notably, the simulations indicate a strong resilience against the collisionless $\nabla n$-driven TEM as a result of the quasi-omnigenous magnetic geometry of W7-X, its nearly max-$J$ property and the low experimental $\eta_e$ and $\eta_i$~\cite{Proll12}. Interestingly, MTMs remain unstable even at low $a/L_n$, and this contrasts with a subset of ECRH plasmas with flat density profiles where ITG is expected to prevail~\cite{Beurskens2021}. To examine this discrepancy, we use as reference the ECRH discharge \#20180920.017 which exhibits characteristics of ITG dominance both in the experiment and in flux-matched gyrokinetic simulations~\cite{Carralero21,AgapitoFernando25}. At $\rho=0.4$, its normalized collision frequency $\nu_c=5.6\times10^{-4}$ ($n_e\approx6\times10^{19}$~m$^{-3}$ and $T_e\approx1.4$ keV) is roughly one fourth of the normalized collision frequency of the NBI-heated discharge analyzed ($\nu_c = 2.2 \times 10^{-3}$, with $n_e=7.9\times10^{19}$~m$^{-3}$ and $T_e=0.79$ keV). This is an important parameter as a very low value of collisionality can stabilize MTMs, as shown in Fig.~\ref{F1_lin_characterization}d. Indeed, repeating the linear scan with this ECRH-$\nu_c$, ITG modes appear at flat density profiles, MTMs appear at moderate $a/L_n$ and the TPMs described earlier become dominant at high $a/L_n$, as shown by the dashed line in Fig.~\ref{F4_shat_omTi_omTe_omn_scans}d.

Two qualitative conclusions emerge from these results. First, the MTM growth rate is mostly insensitive to density gradient, agreeing with MTMs being primarily driven by electron temperature gradient. Second, the transition observed numerically from the reference ITG-dominated ECRH plasma to MTM dominance in NBI plasmas cannot be explained by changes in density gradient alone, but it also requires a higher collisionality with respect to the reference ECRH case. However, quantitatively comparing multiple ECRH and NBI-heated plasmas and studying the detailed ITG-MTM transition are outside of the scope of this work. 

Other instabilities proposed for high-$\nabla n$ regimes in W7-X -- such as the iTEM~\cite{Xanthopoulos20} and the universal instability (UI)~\cite{Costello23,Proll22,Podavini24,Thienpondt25} -- are absent in our simulations. This discrepancy arises because earlier studies neglected collisionality and $\beta_e$, both of which have strong stabilizing effects: collisionality suppresses the iTEM~\cite{Morren24} (see also Fig.~\ref{F1_lin_characterization}d), while $\beta_e$ stabilizes the UI~\cite{Duff25,Huba82,Hastings82}.

\begin{figure}[t]
\includegraphics{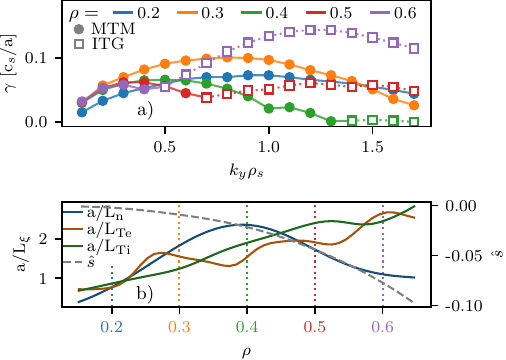}
\caption{\label{F5_xpos_ky_scan} a) Growth rates and unstable modes as a function of $k_y \rho_s$ at different radial positions. MTMs are indicated by circles and solid lines, while ITGs by hollow squares and dotted lines. b) Normalized gradient scale lengths and magnetic shear $\hat{s}$ versus radius. The color of the vertical dotted lines indicate the radial positions of the linear simulations in a).}
\end{figure}

In addition, MTM activity is not localized and remains unstable throughout the radial interval $0.2 \leq \rho \leq 0.6$, as illustrated in Fig.~\ref{F5_xpos_ky_scan}a, which show linear simulations including the corresponding experimental parameters at the radial positions analyzed. MTM is facilitated by the low magnetic shear $\hat{s}$ in the inner plasma, which mitigates MTM damping and the sufficiently large density gradients that stabilize ITG modes, as shown in Fig.~\ref{F5_xpos_ky_scan}b.


\textit{Microtearing mode turbulence.} 
Nonlinear gyrokinetic simulations using the full set of experimental parameters for $\rho=0.4$ yield turbulent fluxes in good agreement with the measured ones: the simulated heat and particle fluxes are $A \langle Q\rangle_{\mathrm{sim}} = 0.84 \pm 0.02$~MW and $A \langle \Gamma\rangle_{\mathrm{sim}} = 4.3 \pm 0.03 \times 10^{19}$~s$^{-1}$, compared to the experimental estimates of $A \langle Q\rangle_{\mathrm{expt}} = 0.6$~MW and $A \langle \Gamma\rangle_{\mathrm{expt}} = 9.9 \pm4.4\times 10^{19}$~s$^{-1}$ (see Figs.~8 and 12 of Ref.~\cite{Ford24}). Here, $\langle \cdot \rangle$ indicates a flux-surface and time average.

\begin{figure}[t]
\includegraphics{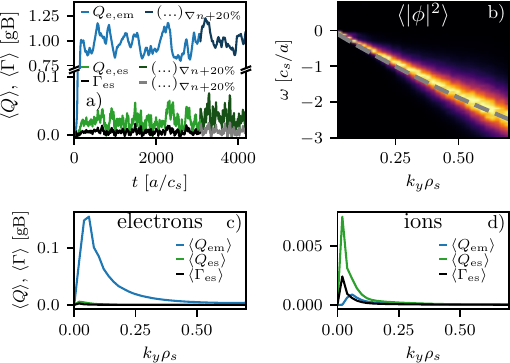}
\caption{\label{F2_timetrace_freq_fluxspec} 
a) Time trace of flux surface-averaged electron heat and particle flux in gyro-Bohm units ($[\Gamma]=n_ec_s(\rho_s/a)^2$ and $[Q]=n_eT_ec_s(\rho_s/a)^2$). Particle flux, electrostatic and electromagnetic heat flux after increasing $a/L_n$ by 20\% are shown in gray, dark green and dark blue, respectively. b) Frequency spectrum of electrostatic potential as a function of wavenumber, with the dashed line indicating the real frequency obtained in linear simulations (Fig.~\ref{F1_lin_characterization}a). 
c) and d) heat and particle flux spectra of electrons and ions, respectively.
}
\end{figure}

The time evolution of the turbulent heat fluxes, shown in Fig.~\ref{F2_timetrace_freq_fluxspec}a, reveals that the electron electromagnetic heat flux dominates over the electrostatic contributions -- a well-established signature of MTM turbulence~\cite{Kotschenreuther19}. ITG turbulence is absent due to a combination of the low $\eta_i$ and moderate $\nu_c$, consistent with the linear scans in Figs.~\ref{F4_shat_omTi_omTe_omn_scans}c and~\ref{F4_shat_omTi_omTe_omn_scans}d. Turbulence generated by the electron temperature gradient (ETG) mode is considered negligible as this mode is stable in linear simulations (not shown), in line with the low $\eta_e = 0.65$. Both branches of the TEM -- those driven by $\nabla T_e$ and $\nabla n$ -- are absent. The background $E \times B$ shear has minimal impact on the time-averaged electromagnetic flux, with $\hat{\gamma}_{E \times B} = 0.037$ $[c_s/a]$ (not shown), consistent with standard tokamak simulations where MTM growth rates are comparable to the shearing rate~\cite{Doerk11,Jian19}. 

The heat and particle flux spectra shown in Figs.~\ref{F2_timetrace_freq_fluxspec}c and~\ref{F2_timetrace_freq_fluxspec}d confirm that the electrostatic ion and electron contributions to the overall heat flux are small. While MTMs typically generate negligible particle transport in tokamaks~\cite{Doerk11}, in this particular case they appear to drive all the particle flux, which is mostly electrostatic. This can be observed in the frequency spectrum (Fig.~\ref{F2_timetrace_freq_fluxspec}b), where the electrostatic potential matches the MTM frequencies identified in linear simulations (Fig.~\ref{F1_lin_characterization}a), and no other dominant or subdominant modes at relevant wavenumbers for the particle flux are found.

\textit{Discussion.} MTM turbulence is characterized by $Q_e \gg Q_i$.
This situation is experimentally plausible in NBI-heated plasmas because, as explained in Ref.~\cite{Ford24}, it is within the limits of diagnostic uncertainty that the turbulent ion heat flux $Q_i$ is zero at the analyzed radius of this NBI-heated plasma.

Two other experimental observations support the existence of MTM turbulence in high-density-gradient regimes. First, in a set of discharges with $a/L_n>1$, the ion heat flux channel is negligible and the electron contribution dominates~\cite{Wappl25}. Second, the total turbulent heat flux for discharges with $a/L_n > 1$ (including this one), is low compared to cases with $a/L_n<1$ and is not exacerbated by the density gradient~\cite{Wappl25,Ford24}, suggesting that the main driving instability is insensitive to this parameter. This behavior is explained if MTMs dominate: the turbulent heat flux is primarily carried by electrons ($Q_{tot} \approx Q_e$), and is unaffected by the density gradient, as observed in dark blue in Fig.~\ref{F2_timetrace_freq_fluxspec}, where $a/L_n$ was increased by 20\% and the heat flux was not substantially affected. A similar behavior has also been observed in MTM present in tokamak pedestals, as illustrated in Fig.~17 of Ref.~\cite{Kotschenreuther19}. Another argument is that the heat flux follows a quasilinear scaling $Q \sim \gamma / \langle k_\perp \rangle^2$~\cite{Dannert2005,Giacomin25}, provided no subdominant modes are active (as supported in this case by their absence in the spectrum in Fig.~\ref{F2_timetrace_freq_fluxspec}b). As shown in Fig.~\ref{F4_shat_omTi_omTe_omn_scans}d, $\gamma$ remains nearly constant (or even slightly decreases) as $a/L_n$ increases, and this verifies the decoupling between the density gradient and heat flux. This decoupling stands in stark contrast to density-gradient-driven modes such as the $\nabla n$-driven TEM branch, iTEM, or UI, and reflects the fact that MTMs are primarily driven by the electron temperature gradient. 

To further exclude the presence of the iTEM and UI in this case, we performed nonlinear simulations keeping all physical parameters fixed except for $\nu_c$, set to zero, and $\beta_e$, reduced to $10^{-5}$. In these limits, the TPM resembling the iTEM and the UI modes became dominant, respectively. The numerical parameters for the collisionless and the electrostatic cases had the grid resolution and box sizes: $(n_x, n_{k_y}, n_z, n_{v_\parallel}, n_\mu) = \{(128,64,192,32,12); (256,32,384,32,12)\}$, and 
$(l_x, l_y, \mathrm{npol}) = \{(128\rho_s, 157\rho_s, 2.05); (384\rho_s, 209.44\rho_s, 4)\}$, applying the generalized and twist-and-shift parallel boundary conditions after $\mathrm{npol}$ poloidal turns, respectively. In both cases, the total turbulent flux-surface and time-averaged heat and particle fluxes were several orders of magnitude larger than those of the experiment: $A \langle Q_{\nu_c=0}\rangle = 6.7$~MW and $A \langle \Gamma_{\nu_c=0}\rangle = 1 \times 10^{22}~\mathrm{s}^{-1}$ for the collisionless case, and $A \langle Q_{\beta_e=10^{-5}}\rangle = 32$~MW and $A \langle \Gamma_{\beta_e=10^{-5}}\rangle = 4.2 \times 10^{22}~\mathrm{s}^{-1}$ for the electrostatic case. These results rule out iTEM and UI, leaving MTM as the only mechanism consistent with the observed fluxes.

As this discharge exhibits a steep density gradient and MTM‑driven turbulence yields small ion heat flux -- both characteristics of high‑performance plasmas~\cite{Bozhenkov2020}, also included in the $a/L_n>1$ database of Ref.~\cite{Wappl25} -- MTMs may contribute to the heat flux in high-performance cases. Their detailed analysis is left for future work.


\textit{Conclusions.} This Letter demonstrates -- for the first time -- the presence of microtearing mode (MTM) turbulence in the W7-X stellarator. The existence of MTMs in the analyzed NBI-heated scenario can be explained by a combination of multiple factors, namely the sufficiently large density gradient, moderate temperature gradients, and max-$J$ -- which inhibit the appearance of ITG and TEM turbulence. The emergence of MTMs is enabled by the moderate collisionality and the low magnetic shear of W7-X, which reduces its stabilization due to low magnetic field line bending. The simulations presented here reproduce the experimental heat and particle fluxes, with collisionality and electromagnetic effects playing a crucial role. MTM‐driven turbulence yields $Q_e\gg Q_i$, a result that lies within the experimental error bars in the analyzed NBI‑heated discharge. In addition, the heat flux associated with this mode is electron-dominated and is insensitive to the density gradient, in agreement with W7-X experimental observations on plasmas with $a/L_n>1$. 

Although a study of MTMs in reactor-relevant scenarios is beyond the scope of this work, a future stellarator power plant could possibly benefit from the low turbulent transport driven by MTMs (compared to the ITG-driven one) as current designs feature low magnetic shear, max-$J$, and pellet injection~\cite{Lion25,Hegna25}. The edge region is especially favorable, where the comparatively low $T_e$ yields a sufficiently high collision frequency, enabling MTMs, while the steep density gradient produced by local pellet ablation (whose location empirically scales as $T_e^{-5/9}$~\cite{Baylor1997}) stabilizes ITG.


\begin{acknowledgments}
\textit{Acknowledgments.} The authors would like to thank T. Görler, D. Told, A. Di Siena, C. Angioni, F. Sheffield, and P. Costello. The simulations were performed in the HPC Clusters Raven, Viper at the Max Planck Computing and Data Facility, and Leonardo from Cineca HPC. This work has been carried out within the framework of the EUROfusion Consortium, funded by the European Union via the Euratom Research and Training Programme (Grant Agreement No 101052200 - EUROfusion). Views and opinions expressed are however those of the author(s) only and do not necessarily reflect those of the European Union or the European Commission. Neither the European Union nor the European Commission can be held responsible for them.
\end{acknowledgments}

\bibliography{bibliography}

\end{document}